\documentclass[12pt,preprint]{aastex}

\lefthead{HAN \& GAUDI} 
\righthead{DOUBLE-PEAK LENSING EVENTS}

\begin{document}
\title{A Characteristic Planetary Feature in Double-Peaked, High-Magnification 
Microlensing Events}

\author{
Cheongho Han\altaffilmark{1}
and
B. Scott Gaudi\altaffilmark{2}
}

\altaffiltext{1}{Program of Brain Korea 21, Physics Department,
Chungbuk National University, Cheongju 361-763, Korea;
cheongho@astroph.chungbuk.ac.kr}

\altaffiltext{2}{Department of Astronomy, Ohio State University, 
140 West 18th Avenue, Columbus, OH 43210-1173;
gaudi@astronomy.ohio-state.edu}



\begin{abstract}
A significant fraction of microlensing planets have been discovered in
high-magnification events, and a significant fraction of these events
exhibit a double-peak structure at their peak.  However, very wide or
very close binaries can also produce double-peaked high-magnification
events, with the same gross properties as those produced by
planets. Traditionally, distinguishing between these two
interpretations has relied upon detailed modeling, which is both
time-consuming and generally does not provide insight into the
observable properties that allow discrimination between these two
classes of models.  We study the morphologies of these
two classes of double-peaked high-magnification events, and identify
a simple diagnostic that can be used to immediately distinguish
between perturbations caused by planetary and binary companions,
without detailed modeling. This diagnostic is based on the difference
in the shape of the intra-peak region of the light curves. The shape
is smooth and concave for binary lensing, while it tends to be either
boxy or convex for planetary lensing. In planetary lensing this
intra-peak morphology is due to the small, weak cusp of the planetary
central caustic located between the two stronger cusps. We apply
this diagnostic to five observed double-peaked high-magnification
events to infer their underlying nature.  A corollary of our study is
that good coverage of the intra-peak region of double-peaked
high-magnification events is likely to be important for their unique
interpretation.

\end{abstract}




\section{Introduction}

Microlensing has emerged as an important method of discovering
extrasolar planets.  Since the first discovery in 2004, six
microlensing planets have been reported \citep{bond04, udalski05,
beaulieu06, gould06, gaudi07}.  The detection rate is rapidly
increasing and six additional planet candidates were detected during
the 2007 season alone \citep{gould08}.  In contrast to the radial velocity
and transit methods, which are most sensitive to planets that orbit
close to their parent star, the sensitivity of the microlensing method
peaks in the cool, outer regions of planetary systems beyond the `snow
line' (\citealt{gould92}, see also \citealt{gaudi08}).  Furthermore,
the sensitivity of the microlensing method extends to very low-mass
planets \citep{bennett96}.  Thus microlensing is sensitive to planets
with physical properties that are very different from those
discovered by other methods, and the sample of microlensing planets
includes notable planets such as the most distant, the coldest, and
the lowest-mass planets detected to date.  In addition, the recently
reported multiple-planet system OGLE-2006-BLG-109Lb,c \citep{gaudi07}
is also noteworthy in that the planet masses and locations relative to
the snow line are similar to those of Jupiter and Saturn.

Microlensing planet searches are currently conducted using a
combination of survey and follow-up observations.  The primary
microlensing events, caused by stars in the Galactic bulge or
foreground disk, are found by survey observations \citep{soszynski01,
bond01}, which maximize the event rate by monitoring a large area of
the Galactic bulge on a roughly nightly basis.  These data are
analyzed real time, thereby making it possible to issue alerts of
ongoing events in the early stage of lensing magnification.  Follow-up
observations \citep{yoo04, cassan04} are focused on these alerted
events in order to detect the short-lived perturbations to the light
curves of the host stars that are the signals of planetary companions
to these stars.  However, the limited number of telescopes available
for follow-up restricts the number of events that can followed at any
given time.  Thus priority is given to those events which will
maximize the planetary detection probability.  Currently, the highest
priority is given to high-magnification events.  There are several
reasons for this.  First, these events have high intrinsic planet
detection efficiency because the source trajectories of these events
always pass close to the perturbation region around the central
caustic induced by the planet \citep{griest98}.  Second, follow-up
observations can be prepared in advance because the time of
perturbation typically occurs near the peak of the event, which can be
predicted reasonably well from data taken on the rise to the peak.  
Third, the enhanced brightness of the
highly-magnified source near the event peak allows for precision
photometry, which is essential for the proper characterization of the planetary
perturbation. In addition, these bright event peaks can be observed using
small-aperture telescopes, which are much more numerous, thus enabling
continuous and frequent monitoring, which is also essential for proper
characterization.  As a result, four
(OGLE-2005-BLG-071Lb, OGLE-2005-BLG-169Lb, and OGLE-2006-BLG-109Lb,c) of the six reported
microlensing planets were detected through the channel of
high-magnification events.

A common morphology of perturbed high-magnification events is a
double-horned, or double-peaked, structure at the peak of the light
curve.  A double-peaked morphology at the peak of a high-magnification
event can be produced in two very different ways.  The first is when
the source approaches the blunt, back end of the asymmetric, wedge-shaped, central caustic of a
planetary companion at an angle of $\sim 90^\circ$ from the
planet/star axis.  The second arises when the source approaches the
symmetric astroid-shaped caustic of a very wide or very close binary
at an angle of $\sim 45^\circ$ from binary axis.  The light curves of
these two types of events are approximately degenerate in the sense
that one can find a value of the shear (for a wide binary lens) or
quadrupole moment (for a close binary lens) such that the peak heights
and time between the peaks are roughly the same as for the planetary
case.  Fortunately, as has been demonstrated empirically
\citep{albrow02}, they are not perfectly degenerate and thus with
sufficient data quality and quantity it is possible to determine
whether a light curve is due to a planet or a binary.  Distinguishing
between the planetary or close/wide binary interpretations of an
observed double-peaked high-magnification event has heretofore
required detailed modeling (e.g., \citealt{albrow02}).  

In this paper, we study the morphology of double-peaked high
magnification events, and identify a diagnostic feature of the
intra-peak region of these light curves that can be used to immediately
distinguish between the planetary and binary interpretations.
Specifically, the intra-peak region of planetary events is typically
boxy or convex, whereas it is smooth and concave for close or wide
binary lenses.  We provide the physical basis for this difference in
the morphology, which is related to the existence of a third, weak
cusp in the central caustic of planetary lens, which is absent in the
close/wide binary case.  While detailed modeling of observed events is
ultimately required to derive precise values of the underlying physical
parameters, this approach is time-consuming.  This diagnostic can be
used to quickly identify those events that are most likely caused by
planetary companions, and so permit efficient allocation of limited
modeling resources.  Furthermore, our study provides some insight
into the kinds of observations that are needed to discriminate between
these two classes of models.  This can aid in the planning of
observations, and if this diagnostic is applied to events
real-time, can inform decisions about which events to follow
given limited
observational resources.

\section{Lensing Properties}

For a binary lens, the mapping between the lens plane and source
plane can be expressed as
\begin{equation}
\zeta = z - {m_1/M\over \bar{z}-\bar{z}_{L,1}} -
{m_2/M\over \bar{z}-\bar{z}_{L,2}},
\label{eq1}
\end{equation}
where $\zeta=\xi+ i\eta$, $z_{L,j}=z_{L,j}+iy_{L,j}$, and $z=x+iy$ 
are the complex angular positions of the source, lens, and image, 
respectively, $\bar{z}$ denotes the complex conjugate of $z$, $m_j$ 
are the masses of the individual lens components, and $M=m_1+m_2$ is 
the total mass \citep{witt90}.  Here all angles are normalized to 
the Einstein radius corresponding to the total mass of the lens system,
\begin{equation}
\theta_{\rm E}=\left[ {4GM\over c^2} \left({1\over D_L}-{1\over D_S}
\right)\right]^{1/2},
\label{eq2}
\end{equation}
where $D_L$ and $D_S$ are the distances to the lens and source, 
respectively.  For a binary lens, there exist three or five 
images depending on the source position with respect to the positions 
of the lens components.  The magnification of each image is the ratio 
between the areas of the image and source.  This corresponds to the 
reciprocal of the determinant of the Jacobian of the lens mapping
evaluated at the image position, i.e.\
\begin{equation}
A_i=\left\vert 
\left( 
1-{\partial\zeta\over \partial\bar{z}}
{\overline{\partial\zeta}\over \partial\bar{z}}\right)_{z=z_i}^{-1}
\right\vert.
\label{eq3}
\end{equation}
Then, the total magnification is the sum of the magnifications of the
individual images, $A=\sum_i A_i$.

An important characteristic of binary lensing is the existence of
caustics.  They represent the set of source positions at which the
magnification of a point source becomes infinite (i.e.\ where the
determinant of the Jacobian is zero), and mark the boundaries of the
region in the source position where the number images differs by two.
For a binary lens, the caustics form one, two, or three closed curves,
interior to which there are five images.  Each set of caustics is
composed of smooth, concave curves which are fold singularities, which
meet at points which are higher-order, cusp singularities.

\begin{figure}[t]
\epsscale{0.8}
\plotone{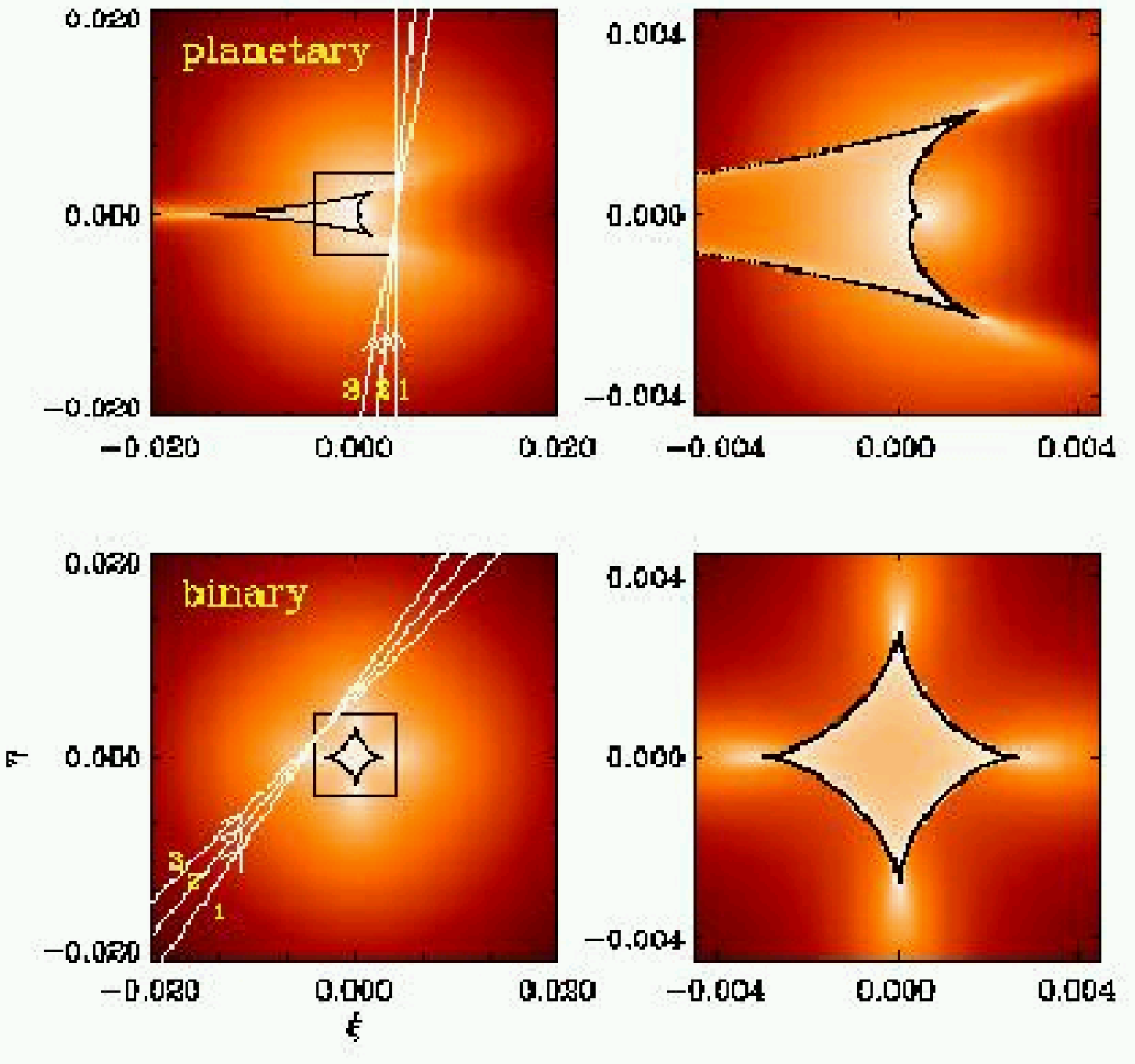}
\caption{\label{fig:one}
Magnification map as a function of the angular position 
of the source for a planet/star (upper panels) and a wide
binary-lens (lower panels) system. The map for the planetary system
shows a small region around the planet-hosting star.  Angles are
in units of the Einstein radius of the planet-hosting star. 
The planet is located to the left
with a separation corresponding to 1.3 times of the star's Einstein
radius and the planet/star mass ratio is $10^{-3}$. The map of the
binary system shows the region in the vicinity of the primary 
lens of the two, equal-mass, binary-lens components.
Angles are in units of the angular Einstein radius of the primary lens.
The other, secondary lens component is located to the left with a
separation corresponding to 26.9 times the Einstein radius of one component.  
In all panels, the coordinates are aligned such that $\xi$ and $\eta$ axes are parallel
with and normal to the line connecting the two lens components,
respectively.  The closed figures drawn in thick black curves
represent the caustics.  The grey-scale is drawn such that brighter
tone represents the region of higher magnification.  The straight
lines with arrows show three different source trajectories leading
to the double-peaked light curves presented in Fig.~\ref{fig:two}.
The maps on the right panels show blowups of the regions enclosed
by the boxes in the left panels.
}\end{figure}

\begin{figure}[t]
\epsscale{0.8}
\plotone{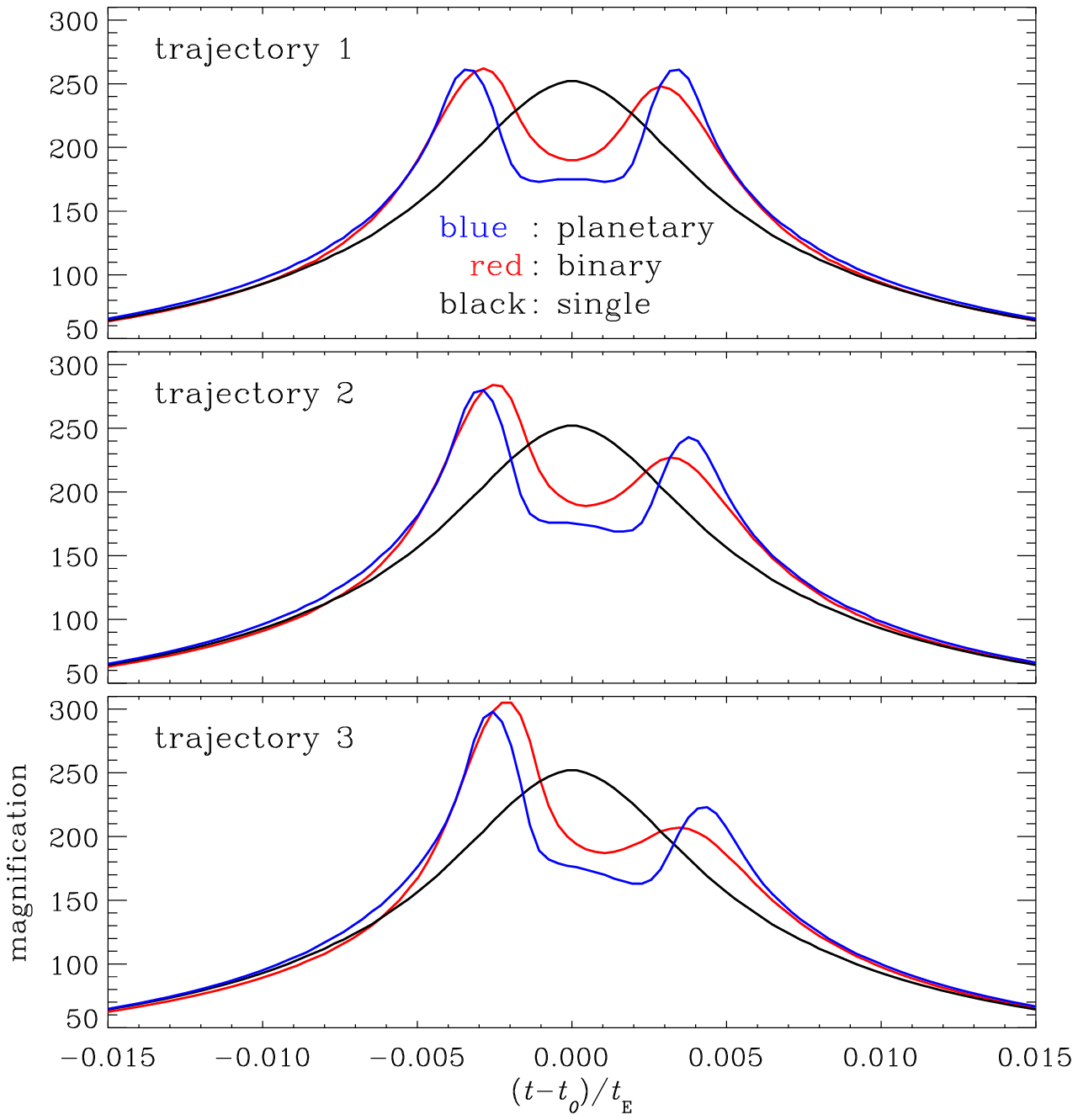}
\caption{\label{fig:two}
Examples of double-peaked high-magnification light curves.  The geometry 
of the lens systems and the source trajectories responsible for the 
light curves are presented in Fig.~\ref{fig:one}.  In each panel, blue and 
red curves are the light curves for the planetary and binary events, 
respectively, whereas the black curve is the light curve for a single lens with
the same mass as the primary lens (i.e., no planetary or binary companion).
The time is relative to $t_0$, the time of closest approach to the origin
of the lens system, and normalized to $t_{\rm E}$, the Einstein time scale
of the primary lens. 
}\end{figure}

\subsection{Planetary Lensing}

Planetary lenses correspond to an extreme case of the binary lens
where the mass of one of the lenses is much smaller than the other.  
In this case, the lens equation can be rewritten in a somewhat
more intuitive form,
\begin{equation}
\zeta = z -{1\over \bar{z}} - {q\over \bar{z}-\bar{z}_p}.
\label{eq4}
\end{equation}
Here the angular coordinates are centered at the position of the planet-hosting 
star, $z_p$ represents the location of the planet, $q$ is the 
planet/star mass ratio, and the angular positions are now normalized to 
the Einstein radius of the primary mass.  
In the case of $q\ll 1$, the analysis of the lensing 
behavior is amenable to a perturbative approach, which yields considerable
insight into the behavior of the caustics and light curves as a function of
the planetary parameters \citep{dominik99,bozza99, 
asada02, an05}.

For the planetary case, unless the $|z_p| \sim 1$, there exist two sets of disconnected caustics.  
One set, which can consist of one or two closed caustic curves, is located away from the host star.
This set is typically referred to as the planetary
caustic (or caustics).  The location of the planetary caustic relative to the source trajectory
depends on the separation between the planet and star, as well as
the angle between the planet/star axis and the direction of motion of the source.  Thus perturbations
due to planetary caustics are not predictable.  

In contrast to the planetary caustic or caustics, the other caustic is always located 
close to the host star, and so is known as the central caustic. Thus central
caustic perturbations always occur at the peak of high-magnification events. 
The central caustic has a
wedge-like shape with {\it four} cusps (see Figure~\ref{fig:one}).  One cusp is located
on the star-planet axis and corresponds to the point of the wedge.  This cusp is
strong, in the sense that light curves from source trajectories that pass reasonably close
to such a cusp will exhibit strong deviations from the single-lens
expectation.   Two of the cusps are located  off the axis on the opposite
side of the caustic, and define the `blunt' end of wedge-shaped caustic.
These two cusps are also strong.  Between these cusps
is region of significant demagnification relative to the single-lens expectation.
The fourth cusp, which is located between these cusps on the planet-star axis, is weak,
in the sense that it creates relatively weak positive deviations.  See Figure~\ref{fig:one}.  
Because of this wedge-shaped geometry, central caustic perturbations typically
occur when the source passes close to either the point or the blunt end of the wedge on a 
trajectory that is approximately perpendicular to the 
planet/star axis.  In the case that it passes the blunt end, the resulting light curve is
double peaked.   In fact, both of the planetary microlensing
events arising from central caustic perturbations have been double-peaked, suggesting
that this class of events might be quite common.  

The size of the central caustic depends on both the star-planet 
separation and planet/star mass ratio.  When the size is measured as 
the separation between the two on-axis cusps, it is related to the 
separation and mass ratio by
\begin{equation}
\Delta\xi_c \simeq {4q\over (s-s^{-1})^2},
\label{eq5}
\end{equation}
where the separation $s$ is expressed in units of the Einstein radius.  
Unlike the size, which depends on both $s$ and $q$, for $q\ll 1$ the shape of the 
caustic is solely dependent on $s$ and it becomes more elongated as 
$s\rightarrow 1$.  For a given mass ratio, a pair of central caustics 
with separations $s$ and $s^{-1}$ are identical to first order in $q$.
For more details about the properties of central caustics, see \citet{chung05}.

In the upper panel of Figure~\ref{fig:one}, we present the central 
caustic of an example planetary lens system and the magnification 
pattern around the caustic.  The planet has a mass ratio $q=10^{-3}$ 
and it is located on the left side of the host star with a separation 
$s=1.3$.  The coordinates are centered at the location of the host star 
and the axes are aligned such that $\xi$ and $\eta$ axes are parallel 
with and perpendicular to the star-planet axis, respectively.  All angular
positions are normalized in units of the Einstein radius corresponding to the 
mass of planet-hosting star.  The grey-scale is drawn such 
that brighter tone represents the region of higher magnification. The 
straight lines with arrows are example source trajectories producing 
double peaked events where the light curves of the resulting events 
are presented in Figure~\ref{fig:two} (blue curves).  The source 
trajectories have a common impact parameter from the primary star but the 
angles with respect to the star-planet axis (source trajectory angle 
$\alpha$) are different.  For the planetary case, the two peaks have 
a similar height when the source trajectory angle is $\alpha\sim 
90^\circ$ and the difference in heights increases as the angle 
deviates from this angle.  The map in the upper right panel shows 
the blowup of the region enclosed by a box in the left-side map.

\subsection{Wide/Close Binary Lensing}

In the limiting case of a binary lens where the projected separation 
between the lens components is much larger than the Einstein radius
($s\gg 1.0$), the lensing behavior in the vicinity of one of the 
lens components (the primary) can be approximated by a Chang-Refsdal lens
\citep{chang79, chang84, dominik99}, i.e.
\begin{equation}
\hat{\zeta}=\hat{z}-{1\over \hat{z}} + \gamma \hat{z}.
\label{eq6}
\end{equation}
Here the notations with `hat' represent angular scales normalized 
by the Einstein radius of the primary.
The quantity $\gamma$ represents the shear induced by the other 
binary component (companion) and it is related to the lens parameters 
by
\begin{equation}
\gamma = {q\over \hat{s}^2},
\label{eq7}
\end{equation}
where $q=m_2/m_1$ is the companion/primary mass ratio, $m_1$ and 
$m_2$ are the masses of the primary and companion, respectively, 
and $\hat{s}$ is the separation between the primary and 
the companion in units of the 
Einstein radius of the primary, which is related to the
separation in units the Einstein radius of the total mass of the binary by $\hat{s}=(1+q)^{1/2}s$.

The shear exerted by the companion 
results in the formation of a small caustic near
the location of the primary 
lens.  In the Chang-Refsdal limit, the caustic has a shape of hypocycloid with four cusps (an astroid) 
regardless of the binary separation and mass ratio.  Two of the cusps
are located on the binary-lens axis, and the other two are along a line perpendicular
to the axis.  All of these cusps are of equal strength. Thus a source trajectory 
that passes close to the caustic on a trajectory with an orientation of $\sim 45^\circ$
with respect to the binary axis will produce a double-peak event with roughly 
equal peak heights. 
The size of 
the caustic as measured by the separation between the two on-axis cusps is,
\begin{equation}
\Delta\xi_c \simeq 4\gamma,
\label{eq8}
\end{equation}
and thus $\Delta\xi_c \propto q$ and $\Delta\xi_c \propto \hat{s}^{-2}$.  

For a close binary with $s\ll 1.0$, the caustic and the magnification
pattern around it are approximately identical to those of the wide
binary with a separation of $s^{-1}$ except that the caustic is 
located at the center of mass of the binary.  In this case, the
size of the caustic is set by the quadrupole moment of the binary.

In the lower left panels of Figure~\ref{fig:one}, we present the 
magnification pattern in the vicinity of the primary of a wide binary 
lens. The companion is located on the left side with a separation of
$\hat{s}=26.9$ and the companion/primary mass ratio is 1.0.  In the 
map, we also mark several example source trajectories resulting in 
double peaked high-magnification events, where the light curves are 
presented in Figure~\ref{fig:two} (red curves).  These 
trajectories have $\alpha\sim 45^\circ$, and
thus the two peaks have a similar height.  The notations are same as those of the 
maps of the planetary lens. 

\section{Difference in Magnification Pattern}

The perturbations from the single-lens form exhibited in double-peaked
high-magnification events can be characterized by three gross
observables: the height of each peak and the time between the two peaks.
Given an observed double-peaked high-magnification event, it is
always possible to find a planetary or binary-lens model that can
reproduce these observables.  In particular, these three observables
can be matched by varying the following three parameters: (1) the angle of the source
trajectory relative to the binary lens axis, which sets the relative
peak heights, (2) the impact parameter from the primary lens\footnote{It
is not strictly true that it is possible to vary the impact parameter
arbitrarily, as this parameter is constrained by the light curve
data away from the peak.  However, for high-magnification events in the usual
highly-blended case, the impact parameter is poorly constrained, and thus
our discussion is approximately correct.}, which sets
the average height of the two peaks, and (3) either the shear (in the
case of a wide binary), the quadrupole moment (in the case of a
close binary), or the parameter combination $q (s-s^{-1})^{-2}$ (in
the case of the planetary lens), which set the size of the caustic and so
the time between the peaks.   

Although it is possible to match these three gross observables with either a planetary or
wide/close binary lens, the morphology of these two classes of
double-peaked high-magnification events are not identical, as
illustrated in Figure~\ref{fig:two}.  The most noticeable difference
arises in the shape of the intra-peak, trough region.  In the case of
a close/wide binary lens, this region has a smooth, rounded, concave
shape. On the other hand, in the case of the planetary lens curve,
this region has a boxy, slightly convex morphology.

The morphology of the intra-peak trough region in the planetary case
is caused by the existence of the fourth cusp in the central caustic
located in between the two stronger cusps (see the blowup of the
planetary central caustic in the upper right panel of
Fig.~\ref{fig:one}).  The general pattern of magnification around a
caustic is a lobe of positive perturbation in the region
immediately surrounding the cusp, flanked on the side of the
cusp by a more extended region of relative demagnification.  
For the planetary case, the weak fourth cusp
results in either a boxy intra-trough region, or even a slight
convexity, caused by the lobe of positive perturbation associated with
the weak cusp `filling in' the trough created by the two neighboring,
stronger cusps.  Since the cusp is weak, this bump is generally weak,
but its effect on the morphology of the light curve is generally not
negligible. For close/wide binary lenses, on the other hand, there is
no weak middle cusp.  Thus the intra-peak morphology for a binary lens
caustic is characterized by the double peaks that occur when the
source approaches the strong cusps of the astroid Chang-Refsdal
caustic, and a smooth, concave intra-peak trough that occurs as the
source passes through the negative perturbation region between the two
cusps.

\begin{figure}[t]
\epsscale{0.8}
\plotone{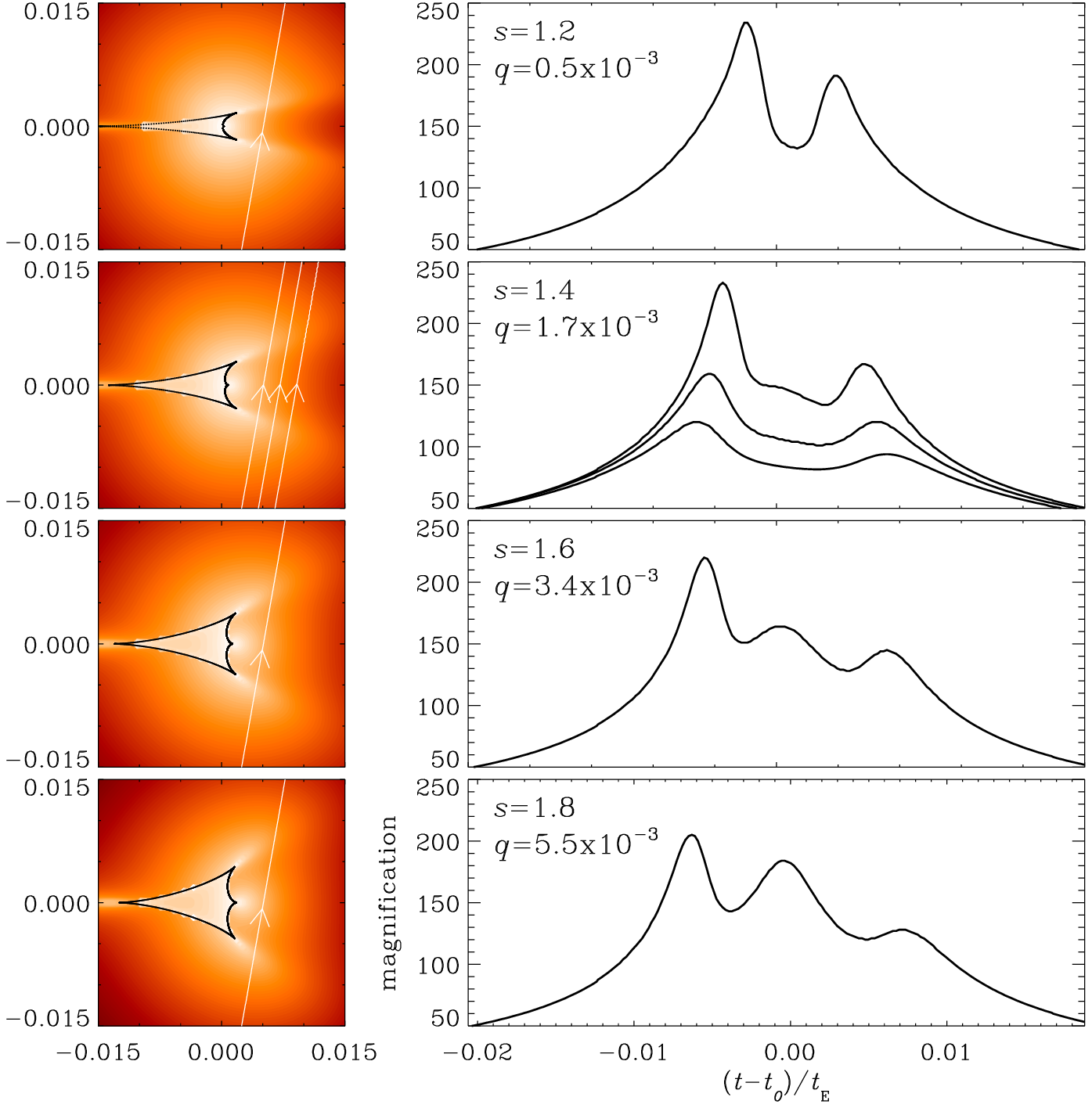}
\caption{\label{fig:three}
Variation of the intra-peak morphology of double-peaked high-magnification
planetary microlensing events.  The left panels show the magnification 
patterns around central caustics of various shapes.  In each panel,
the white lines indicate example trajectories, with the
corresponding light curves shown in the right panels.  In the second row,
several trajectories are shown to illustrate the effect
the varying the impact parameter of the source 
trajectory on the intra-peak morphology.  
Notations for the magnification pattern maps are same as 
in Fig.~\ref{fig:one}.  The values marked in each panel represent the 
star-planet separation in units of the Einstein radius ($s$) and the 
planet/star mass ratio ($q$).
}\end{figure}

The impact of the weak middle cusp on the planetary lensing light
curve morphology varies depending on the underlying parameters.  Two
factors affect the shape of the feature.  The first is the
overall shape of the central caustic.  This can be seen in
Figure~\ref{fig:three}, where we present magnification patterns around
central caustics of various shapes, as well as example light curves.  In order
to isolate the effect of the caustic shape on the morphology, we have
adjusted the parameters so that the caustics have a similar size
(i.e., so that the parameter combination $q (s-s^{-1})^{-2}$ is
constant).  The shape of the caustic depends on the star-planet
separation $s$, and thus the shape of the intra-peak feature depends
on the planetary separation.  The variation is such that the cusp
responsible for the intra-peak feature is stronger and thus the
intra-peak feature is more prominent for planet located further from
the Einstein radius of the central star.  The other factor that
affects the shape of the intra-peak feature is the impact parameter of
the source trajectory.  As the source trajectory passes closer to the
weak cusp, the resulting feature becomes more obvious.  The variation
of the intra-peak feature with the impact parameter is shown in
the second panels of Figure~\ref{fig:three}.

We note that although the intra-peak region shows the most obvious
morphological differences between the close/wide binary and planetary
light curves, there are other, somewhat more subtle differences which
may also be used to distinguish between these two interpretations.  In
particular, the detailed shapes of the peaks appear to differ.  In the
planetary case, the lobes of high magnification due to the strong
cusps are asymmetric about the symmetry axes of the cusps, whereas for
the binary-lens case the magnification patterns near the cusps are
nearly perfectly symmetric about the symmetry axes.
Thus, even in the case of a poorly sampled intra-peak region, or when
$s \rightarrow 1$ in the planetary lens case such that the middle
cusp is very weak, detailed information about the shape of the peaks
will allow one to distinguish between the two models.

An important corollary to our study is that one expects that
high-magnification, double-peaked events in which the intra-peak region
is poorly sampled to be subject to more severe degeneracies, such
that the span of allowed models is larger.  Thus good coverage of the intra-peak
region is likely
to be important for the unique interpretation of these events.

\section{Application to Observed Events}

The diagnostic feature we have identified is useful as it 
can be used to distinguish between planetary and close/wide binary
interpretations of observed events, without the need for detailed
fitting. Thus it can be used to quickly identify those events
which are most likely due to planetary companions. Motivated by this,
we apply our diagnostic to five high-magnification, double-peaked events
with reasonably well-covered peaks
\footnote{We do not consider events that are technically
double-peaked, high magnification events, but are clearly not produced
by the two classes of models we have considered in this paper. An
example is MOA-2002-BLG-33, which has a maximum magnification of
$\sim 450$ and exhibits a double-peaked morphology, but is clearly
caused by a geometry in which the source trajectory crosses a caustic with a size of order 
the source size \citep{abe03}.}. For
two of these events, MACHO 99-BLG-47 \citep{albrow02} and
OGLE-2005-BLG-071 \citep{udalski05,dong08}, detailed modeling has
already been done.  Three additional events were observed during 2007
lensing season, for which detailed modeling has not been reported.
These events are OGLE-2007-BLG-349/MOA-2007-BLG-379, OGLE-2007-BLG-514,
and OGLE-2007-BLG-137/MOA-2007-BLG-091.
\begin{enumerate}
\item {MACHO 99-BLG-47}: This is the first published high-magnification 
event with two well-resolved peaks.  The trough between the two peaks 
of this event exhibits a smooth concave shape suggesting that the lens is a 
wide/close binary, and not a planetary system.  The large difference between 
the heights of the two peaks implies that the source trajectory angle is 
considerably different from $45^\circ$.  This diagnostic matches the 
results from the detailed analysis of the event conducted by \citet{albrow02}.  
In this model, the source trajectory angle was estimated to be $\alpha\sim 
25^\circ$.
\item {OGLE-2005-BLG-071}: The intra-peak trough of this event exhibits 
a prominent convex feature, implying that the perturbation is caused by 
a planet.  In addition, the two peaks are of almost equal height, implying
that the source passes nearly perpendicular to the binary axis.  This 
diagnostic matches the results obtained from the detailed modeling 
conducted by \citet{udalski05} and \citet{dong08}.
\item {OGLE-2007-BLG-349/MOA-2007-BLG-379}: This event was detected 
during 2007 season and its peak was densely covered by follow-up 
observations.  The two peaks have moderately different heights and the 
trough between the peaks has a linear structure.  These characteristics 
are similar to the planetary lensing light curve presented in the middle 
panel of Figure~\ref{fig:two} except with the direction of time 
reversed.  Our diagnostic would therefore indicate that this event 
is caused by a planet with a source trajectory angle somewhat different 
from $\alpha=90^\circ$.
\item {OGLE-2007-BLG-514}: This is another double peaked high-magnification 
event observed during 2007 season.  The intra-peak trough shows a smooth 
concave shape and thus we diagnose that the lens is a either wide or close 
binary and not a planetary system.  The heights of the two peaks are similar 
and thus the source trajectory angle is close to $\alpha=45^\circ$.
\item {OGLE-2007-BLG-137/MOA-2007-BLG-091}: The apparent magnification
of this double-peaked event is only a modest $\sim 10$, although the
true magnification could be substantially higher if it is highly
blended.  The intra-peak trough is not well-covered, but the data
in this region appear to show a relatively sharp change in the slope
of the magnification, perhaps followed by a linear rise.  This
morphology is indicative of a planetary (or at least low mass ratio)
companion.  The heights of the two peaks are quite different, implying
a source trajectory angle significantly different from
$\alpha=90^\circ$, assuming the event is due to a low mass-ratio companion.
\end{enumerate}

\section{Conclusion}

We have investigated the morphology of double-peaked high magnification
events, which can be produced by two very different classes of models:
planetary lenses in which the source trajectory passes close to the
back end of the wedge-shaped central caustic, and very wide or very
close binary lenses in which the source passes close to two of the
cusps of a Chang-Refsdal caustic. From a comparison
of the morphology of the light curves produced by these two classes of
models we have identified a diagnostic that can be used to
immediately distinguish between perturbations produced by a planet
from those produced by a binary companion.  This diagnostic is based
on the difference in the shape of the intra-peak region of the light
curve.  For binary lensing, the shape is smooth and concave, whereas
for planetary lensing, the shape is boxy or convex.  The morphology of
the intra-peak region of planetary lensing events is due to the
existence of a weak cusp located between the two stronger
cusps. Finally, we applied this diagnostic to five observed
double-peaked high-magnification events.

\acknowledgments 

This work was supported by the Science Research Center (SRC) program.  
We would like to thank A. Gould for providing light curves of the events observed 
by the MicroFUN collaboration during the 2007 season.


\end{document}